\documentclass[aps,twocolumn,footinbib]{revtex4-1}
\usepackage{amsmath,amssymb}
\usepackage{bm}
\usepackage{epsfig}
\usepackage{natbib}
\usepackage{hyperref}
\usepackage{color}
\usepackage{graphicx}

 \hypersetup{pdfstartview={FitH},pdfpagemode={UseNone},
            colorlinks,linkcolor=red, citecolor=blue, urlcolor=blue,
            bookmarks=true, bookmarksopen=true, pdfnewwindow=true}

\begin{document}

\title{Spin and optical properties of silicon vacancies in silicon carbide (a review)}

\author{
S.~A.~Tarasenko$^1$,
A.~V.~Poshakinskiy$^1$,
D.~Simin$^2$,
V.~A.~Soltamov$^1$,
E.~N.~Mokhov$^{1,3}$,
P.~G.~Baranov$^1$,
V.~Dyakonov$^{2,4}$,
G.~V.~Astakhov$^2$
}

\affiliation{
  $^1$\,Ioffe Institute, 194021 St.~Petersburg, Russia\\
  $^2$\,Julius-Maximilian University of W\"{u}rzburg, Experimental Physics VI, 97074 W\"{u}rzburg, Germany\\
	$^3$\,ITMO University, 197101 St. Petersburg, Russia\\
  $^4$\,Bavarian Center for Applied Energy Research (ZAE Bayern), 97074 W\"{u}rzburg, Germany
  }
  

\begin{abstract}

We discuss the fine structure and spin dynamics of spin-3/2 centers associated with silicon vacancies in silicon carbide. 
The centers have optically addressable spin states which makes them highly promising for quantum technologies. 
The fine structure of the spin centers turns out to be highly sensitive to mechanical pressure, external magnetic and electric fields, temperature variation, etc., which can be utilized for efficient room-temperature sensing, particularly by purely optical
means or through the optically detected magnetic resonance. We discuss the experimental achievements in magnetometry and thermometry based on the spin state mixing at level anticrossings in an external magnetic field and the underlying microscopic mechanisms. We also discuss spin fluctuations in an ensemble of vacancies caused by interaction with environment.

\end{abstract}

\maketitle


\section{Introduction}

Silicon carbide (SiC) has been attracting growing attention due to diversity of its polytypes with remarkable and tunable electric and optical properties as well as radiation stability. Of particular interest for fundamental research in optics and applications in sensorics and spintronics are optically addressable color centers associated with the defects of crystal lattice. Such color centers are characterized by spin-dependent optical cycles: their spin states can be selectively initialized and read-out by optical means and, additionally, efficiently manipulated by a radiofrequency (RF) field~\cite{Childress2014}. The color centers discovered in 
SiC~\cite{Vainer:1981vj} can be gathered in two wide classes depending on their spin in the ground state, $S = 1$ or $S= 3/2$. 

Examples of color centers with the integer spin $S=1$ in SiC and, accordingly, with the triplet ground state  include 
silicon-carbon divacancies of the neighbouring positions with covalent molecular bond (known as P6/P7)~\cite{Vainer:1981vj,Baranov2005,Sorman2000,Christle2014} and recently discovered silicon vacancy-nitrogen pairs (V$_{\rm Si}$N$_{\rm C}$)~\cite{Bardeleben2015,Bardeleben2016}. The spin and optical properties of these triplet centers are quite similar to
those of the well-known negatively-charged nitrogen vacancy center (NV$^{-}$) in diamond~\cite{Doherty2013}. In particular, the zero-field spin splitting of the ground state between the levels with the spin projections $m = \pm 1$ and $m = 0$ along the defect axis is in the GHz range as it is for NV$^{-}$ centers in diamond 2.9~GHz. The centers can be optically pumped into the state with the spin projection $m=0$ through a spin-dependent intersystem-crossing pathway~\cite{Baranov2005,Son:2006im,Falk:2013jq}. 

Centers with the half-integer spin $S=3/2$ and, accordingly, the quadruplet ground state are represented in SiC by silicon vacancy-related centers (V$_{\rm Si}$, also denoted as V2 in literature)~\cite{Mizuochi:2002kl,Kraus:2013di}. They are negatively charged silicon vacancies in the paramagnetic state~\cite{Vainer:1981vj,Wimbauer:1997fj,Sorman2000,vonBardeleben:2000jg,Wagner:2000fj,Mizuochi:2003di,Isoya:2008fx}  that, according to the ENDOR experiments~\cite{Soltamov2015}, are noncovalently bonded to neutral carbon vacancies in the non-paramagnetic state located on the adjacent site along the SiC $c$-axis. The ground state of V$_{\rm Si}$ in zero magnetic field is split into two Kramers doublets with the zero-field spin splitting lying in the MHz range~\cite{Orlinski:2003dw}.
Due to bright and stable photoluminescence (PL), moderate value of the zero-field splitting, narrow broadening of spin sublevels,  V$_{\rm Si}$ are proposed for appealing quantum applications \cite{Baranov:2011ib,Riedel:2012jq,Castelletto:2013jj}. They can be coherently manipulated at room temperature~\cite{Soltamov:2012ey,Kraus:2013di,Carter2015,Embley:2017bf} and reveal a long spin coherence exceeding $20 \,$ms at cryogenic temperatures without isotope purification~\cite{Simin:2017iw}.  Together with demonstrated integration of spin centers into SiC photonic cavities~\cite{Calusine:2016hr,Radulaski:2017ic,Bracher:2017il} and narrow zero-phonon line (ZPL) transitions~\cite{Riedel:2012jq}, these observations make the SiC platform very attractive for quantum communication and information processing.   

Furthermore, because the fine structure of V$_{\rm Si}$ is sensitive to external fields, temperature variation, mechanical pressure, etc., these centers can be utilized for quantum sensing and metrology~\cite{Kraus:2013vf,Soykal:2016tk}. Particularly, high-precision vector magnetometry \cite{Simin2015,Lee2015,Niethammer2016}, all-optical magnetometry \cite{Simin2016} and thermometry \cite{Anisimov:2016er} as well as strain sensitivity \cite{Falk:2014fh} with V$_{\rm Si}$ and other vacancy-related defects have been demonstrated. 
The V$_{\rm Si}$ centers can be integrated into SiC nanocrystals~\cite{Muzha:2014th}, and given their non-toxicity  as well as  near infrared emission and excitation spectral ranges~\cite{Hain2014}, where the absorption in tissue is weak, they are potentially suitable for monitoring physical and chemical processes in living organisms. 

The technology of the growth of high-quality monocrystalline SiC has been well
developed~\cite{Lely1955,Tairov1978,Vodakov1979,Kordina1996}. The commonly used methods for the growth of large-size SiC crystals are
the modified Lely method, also known as the Physical Vapor Transport (PVT)~\cite{Lely1955,Tairov1978}, and the High Temperature Chemical Vapor Deposition (HTCVD)~\cite{Kordina1996}. PVT is based on the sublimation of a polycrystalline SiC source, placed in a hot zone of the growth chamber, and the transport of the vapor phase to a SiC substrate, placed in a relatively cool zone of the chamber, on which the crystal growth occurs. More detailed description of the PVT of SiC can be found, e.g., in Refs.~\cite{Tairov1978,Jenny2004}.
HTCVD is a development of the CVD technology with significantly higher growth temperature ($>$1900$^\circ$C) which enables a high-rate growth of epitaxial SiC films. More details about the HTCVD of SiC including the growth mechanisms and the types of reactors
can be found in Refs.~\cite{Kordina1996,Ellison1999}. Both techniques provide high-purity (the concentration of uncontrollable impurities
is less than $10^{15}$\,cm$^{-3}$) large-diameter (up to 150\,mm) SiC wafers, crystals of $n$- or $p$-type conductivity, and
semi-insulating crystals~\cite{Kordina1996,Jenny2004}. All the  mentioned above types of SiC single-crystal substrates are commercially available. Moreover, isotopic engineering of SiC crystals has been successfully demonstrated~\cite{Wagner2002,Ilyin2007,Astakhov2016}. Particularly, shallow donors and deep-level defects have been studied in PVT-grown SiC single crystals enriched with $^{13}$C, $^{28}$Si or $^{29}$Si isotopes.

Using electron \cite{Christle2014,Widmann:2014ve} and neutron \cite{Fuchs2015} irradiation, the vacancy-related defect concentration in SiC can be controlled over 8 orders of magnitude down to single defect level. Recent demonstrations of 3D engineering of defect locations with focused ion beams \cite{Kraus:2017cka,Wang:2017fb} open a path to the fabrication of hybrid quantum chips, for instance, quantum sensors with electrical control \cite{Klimov:2013ua,Cochrane:2016dd} or room-temperature single photon emitters on demand \cite{Fuchs:2013dz,Lohrmann:2015hd}. 

In this review paper, we discuss the fine structure, spin dynamics and optical properties of $V_{\rm Si}$ (V2) spin centers in the most common and technologically available polytype 4H-SiC.  We give an overview of the applications of the spin centers for all-optical measurements of magnetic fields and temperature and discuss the underlying mechanisms and microscopic theory. We also present a theory of spin fluctuations in an ensemble of $V_{\rm Si}$ caused by interaction with environment and show that the correlation function of the luminescence intensity $g^{(2)}$ contains information about the spin fluctuations and the spin relaxation time.

The paper is organized as follows. In Section~\ref{Sec_structure} we describe the structure of spin levels of $V_{\rm Si}$ centers in an external magnetic field. Section~\ref{Sec_optics} is devoted to the optical properties of $V_{\rm Si}$, spin dynamics, and the mechanism of the optical signal formation. In Section~\ref{Sec_metry} the application of $V_{\rm Si}$ for all-optical magnetometry and thermometry is discussed. In Section~\ref{Sec_noise}, we describe briefly the spin noise in an ensemble of $V_{\rm Si}$. 
Section~\ref{Sec_summary} summarizes the paper.  

\section{Spin structure of Si-vacancy-related centers}\label{Sec_structure}

The fine structure of a $V_{\rm Si}$ spin-3/2 center is determined by the real atomic arrangement which is described by the $C_{3v}$ point group. Due to the presence of a preferable axis parallel to the crystal $c$-axis, the ground state of the spin center in zero magnetic field is split into two Kramers doublets with the spin projections $m = \pm 1/2$ and $\pm 3/2$ along the $c$-axis. 
An external magnetic field $\bm B$ splits the levels $\pm 1/2$ and $\pm 3/2$ further. 

The effective Hamiltonian of a spin-$3/2$ center with the $C_{3v}$ symmetry to first order in the magnetic field 
$\bm B = (\bm B_{\perp}, B_z)$ has the form~\cite{Simin2016}
\begin{align}\label{H}
\mathcal{H}=\mathcal{H}_0+\mathcal{H}_{1\parallel}+\mathcal{H}_{1\perp} \,,
\end{align}
where $\mathcal{H}_0$ is the zero-field Hamiltonian,
\begin{equation}
\mathcal H_0 = D \left( J_z^2 - \dfrac{5}{4} \right) \,,
\end{equation}
$\mathcal H_{1\parallel}$ and $\mathcal H_{1\perp}$ are the Zeeman terms in the magnetic fields $\bm B_{\perp}$ and $B_z$, respectively,
\begin{eqnarray}
  &\mathcal H_{1\parallel} =  \left[ g_{\parallel} J_z +  g_{2\parallel} J_z \left(J_z^2 - \dfrac{5}{4} \right) + g_{3\parallel}\dfrac{J_+^3 - J_-^3 }{4{\rm i}} \right] \mu_B  B_z , \nonumber \\
   &\mathcal H_{1\perp} =   g_{\perp}\mu_B \mathbf{J}_\perp \cdot \mathbf{ B}_\perp +  2g_{2\perp}\mu_B \left\{ \mathbf{J}_\perp \cdot \mathbf{ B}_\perp, J_z^2-\dfrac{3}{4} \right\}
 \nonumber \\ &+ g_{3\perp} \mu_B \dfrac{\{J_+^2,J_z \} B_+ - \{J_-^2,J_z\} B_- }{2{\rm i}} \nonumber \,,
\end{eqnarray}
$J_x,J_y,J_z$ are the spin-$3/2$ operators (the matrix forms of these operators can be found, e.g., in Ref.~\cite{IP_book}),
$\bm J_\perp = (J_x,J_y)$, $J_\pm = J_x \pm {\rm i} J_y$, $\bm B_{\perp} = (B_x,B_y)$, 
$B_\pm = B_x \pm {\rm i} B_y$, $\{A,B\}=(AB+BA)/2$ is the symmetrized product, $z$ is the axis parallel to $c$-axis, $x$ and $y$ are the perpendicular axes with $y$ lying in a mirror reflection plane, and $\mu_B$ is the Bohr magneton. 

In systems of the $C_{3v}$ symmetry,
all six $g$-factors ($g_{\parallel}$, $g_{\perp}$, $g_{2\parallel}$, $g_{2\perp}$, $g_{3\parallel}$, and $g_{3\perp}$) describing the
energy spectrum and the structure of spin sublevels in the magnetic field are generally linearly-independent. The parameters of the effective Hamiltonian can be obtained from the spectra of optically detected magnetic resonance (ODMR) in the static magnetic field~\cite{Simin2016}. In structures with weak spin-orbit coupling, such as SiC, the $g$-factor components $g_{2 \parallel}$, $g_{2 \perp}$, $g_{3 \parallel}$, and $g_{3 \perp}$ are small compared to $g_{\parallel} \approx g_{\perp} \approx 2$. The components $g_{3 \parallel}$ and $g_{3 \perp}$ stems from the trigonal asymmetry of the vacancy. In RF magnetic field, they enable magneto-dipole spin transitions with the spin projection change $\Delta m = \pm 2$ which are forbidden in the axial approximation. The presence of such ``forbidden'' lines is clearly visible in ODMR spectra, see, e.g., Ref.~\cite{Simin2015} and the $\nu_3$ and $\nu_4$ lines in Fig.~\ref{fig_ODMR}a.
The microscopic mechanism of ODMR signal formation is discussed below in Sec.~\ref{Sec_optics}.  

The excited state of the vacancy-related center has a similar spin structure. It can be also described by the effective Hamiltonian~(\ref{H}). However, the parameters of the Hamiltonian are different from those for the ground state. 

Figure~1 shows the energy spectrum of the ground and excited states of a silicon vacancy in the magnetic field directed along the $c$-axis of the crystal. At zero field, the ground and excited spin levels are split into Kramers doublets. For the $V_{\rm Si}$(V2) center in 4H-SiC, the zero-field splittings of the ground and excited states at room temperature are $2D/h \approx 70 \, \mathrm{MHz}$~\cite{Sorman2000,vonBardeleben:2000jg} and $2D'/h \approx 410$\,MHz, respectively. 
In the magnetic field $B_z$, the spin sublevels are split further and shift linearly with the field. At certain magnetic fields, the ground-state sublevel with the spin projection $m = - 3/2$ would cross with the sublevels with the spin projections $m = +1/2$ and $m = -1/2$. In reality, due to a small perpendicular magnetic field component $\bm B_\perp$ caused by a slight misalignment of the external magnetic field or due to unavoidable internal magnetic fields caused, e.g., by hyperfine interaction with surrounding nuclei, the level crossings turn into the level anticrossings denoted in Fig.~1 by GSLAC-1 and GSLAC-2, respectively. Similar crossings, 
ESLAC-1 and ESLAC-2, occur between the excited-state spin sublevels at higher magnetic fields.

\begin{figure}[t]
\includegraphics[width=.99\linewidth]{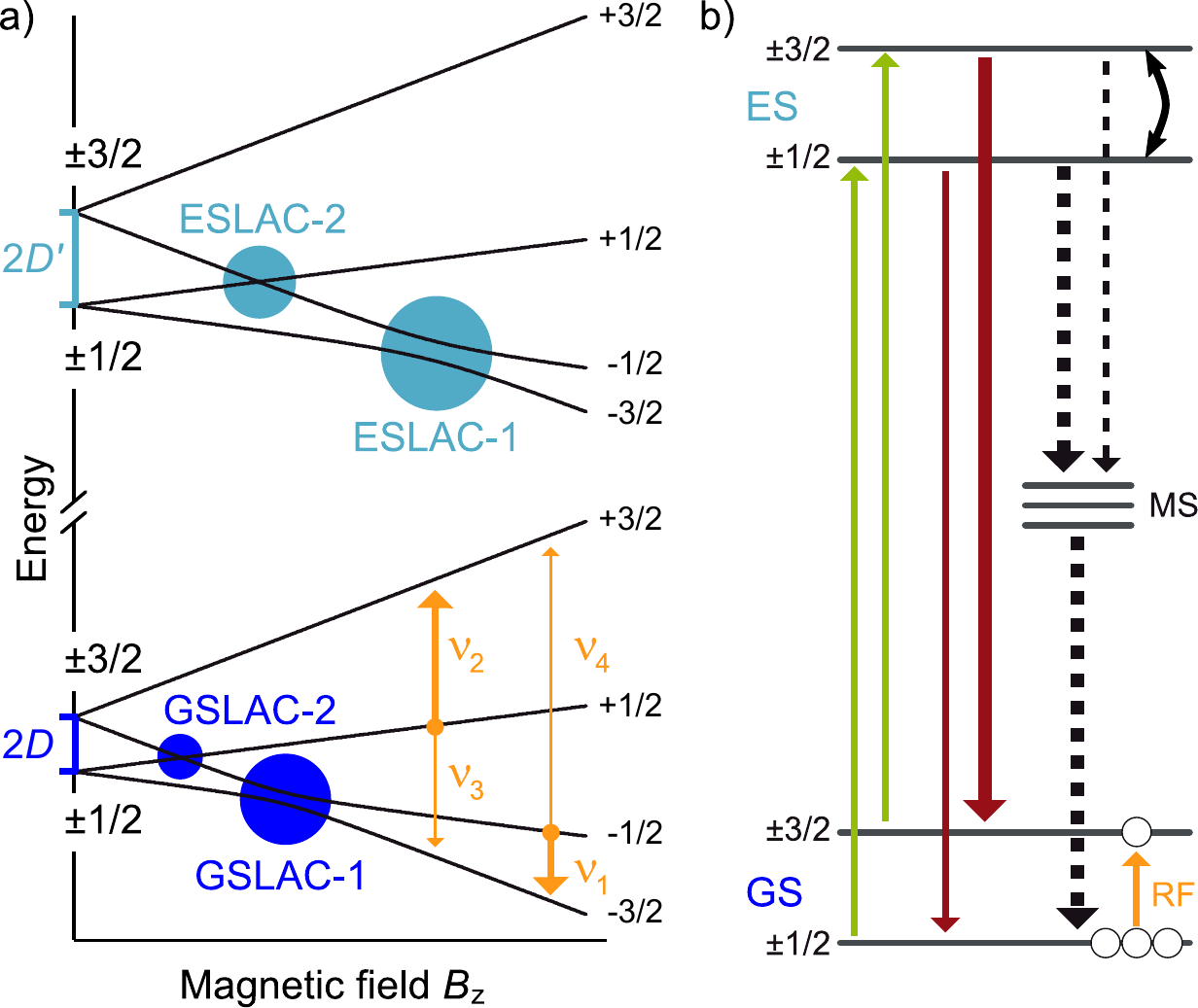}
\caption{(a) The structure of the spin levels of the ground and excited states of a Si-vacancy-related spin center in SiC in an external magnetic field $\bm B \parallel z$. The colored circles visualize the positions of anticrossings and the orange lines labeled $\nu_1 \ldots \nu_4$ denote the spin transitions observed in ODMR spectra. (b) Scheme of the optical initialization and read-out of spin state. Optical pumping leads to a preferential population of the $\pm 1/2$ spin sublevels. The PL intensity is higher when the system is in the $ \pm 3/ 2$ spin states. The populations of spin sublevels can be additionally manipulated by a RF field.}
\label{fig1}
\end{figure}

The described above spin structure of $V_{\rm Si}$ centers can be directly revealed in optical measurements.
This is illustrated below with photoluminescence and ODMR measurements on natural and isotopically purified 4H-SiC crystals. The latter consist of above 99.0\% $^{28}$Si nuclei with zero nuclear spin, in order to elude the hyperfine interaction. The samples contain silicon vacancies of the density $2 \times 10^{14} \, \mathrm{cm^{-3}}$, which were introduced by irradiation with neutrons. To optically address $\mathrm{V_{Si}}$ spin states, a 785-$\mathrm{nm}$ laser diode is used. The PL of $V_{\rm Si}$ centers, which lies in the near-infrared range \cite{Hain2014}, is detected by a Si photodiode through a $900 \,\mathrm{nm}$ long-pass filter. The output signal is locked-in. A static magnetic field $\bm B$ can be applied in an arbitrary direction using a 3D coil arrangement in combination with a permanent magnet. Additionally, to manipulate the $\mathrm{V_{\rm Si}}$ spin states in ODMR measurements, a RF field from a signal generator can be introduced to the sample via a copper stripline. 

Evolution of the ODMR spectrum of V$_{\rm Si}$ (V2) centers in 4H-SiC with the increase of the magnetic field directed along the $c$-axis is shown in Fig.~\ref{fig_ODMR}. At zero field (Fig.~\ref{fig_ODMR}b), two peaks in the ODMR spectrum are visible at the expected spectral positions of the ground-state (GS) and excited-state (ES) zero-field splittings $2D$ and $2D'$, respectively. Due to the short lifetime of the excited state of about $6 \, \mathrm{ns}$~\cite{Fuchs2015}, the latter peak can only be resolved at high RF powers. 

With the increase of the magnetic field strength (Fig.~\ref{fig_ODMR}a), both peaks in the ODMR spectrum are split, as the result of the Zeeman splitting, into four separate lines. Each line is denoted according to the transitions assigned in Fig.~\ref{fig1}a. The magnetic field values, at which the level anticrossings GSLAC-1 and GSLAC-2 occur, are directly visible as the turning points of the $\nu_1$ and the $\nu_3$ lines. While the ODMR signals stemming from the RF-driven spin transitions in the excited state are not directly observed, the magnetic field corresponding to the level anticrossing in the excited state ESLAC-1 can nevertheless be determined from the change in the contrast of the $\nu_1$ line at around $15\,\mathrm{mT}$. The observation of the contrast change in the $\nu_1$ line rather than in the $\nu_2$ line indicates the order of the spin sublevels in the ES, with the $m = \pm 3/2$ state having higher energy than the $m =\pm1/2$ state (positive zero-field splitting 2$D'$ in ES)~\cite{Simin2016,Carter2015}.

\begin{figure}[t]%
\
\includegraphics*[width=.95\linewidth]{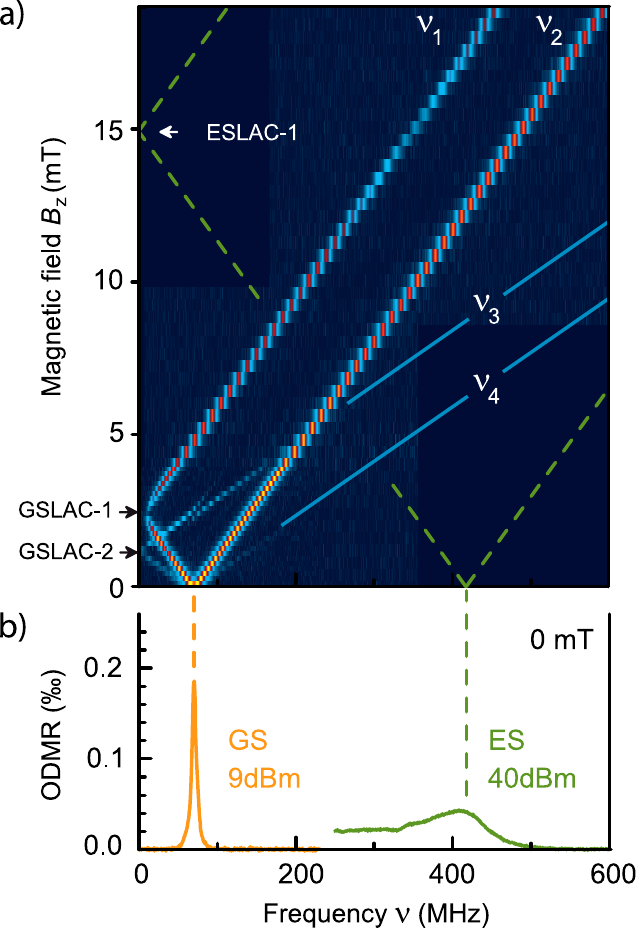}
\caption{(a) Evolution of the ODMR signals from V2 centers in 4H-SiC with the magnetic field along the $c$-axis. The appearing lines are denoted according to the transitions depicted in Fig.~\ref{fig1}. Additionally, solid and dashed lines represent calculated positions of the ODMR peaks in case of a diminishing contrast. (b) ODMR spectrum at zero magnetic field. The two peaks are ascribed to the GS zero field splitting at around $70\,\mathrm{MHz}$ (orange) and the ES zero field splitting at around $410\,\mathrm{MHz}$ (green).}
\label{fig_ODMR}
\end{figure}

\begin{figure}[t]%
\
\includegraphics*[width=.99\linewidth]{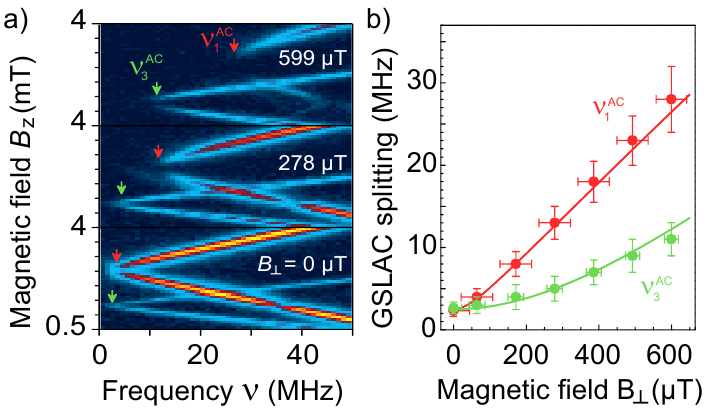}
\caption{(a) Evolution of the $\nu_1$ and $\nu_3$ ODMR lines in the vicinity of GSLACs at different perpendicular field components $B_\perp$. The arrows denote the turning points $\nu^{\mathrm{AC}}_1$ and $\nu^{\mathrm{AC}}_3$, which are direct measures of the energy splittings at GSLAC-1 and GSLAC-2, respectively. (b) The GSLAC splittings $\nu^{\mathrm{AC}}_1$ and $\nu^{\mathrm{AC}}_3$ as a function of $B_\perp$.}
\label{fig_turning}
\end{figure}

The splittings (small energy gaps) between the levels at GSLAC-1 and GSLAC-2 can be determined from the spectral position of the turning points of the $\nu_1$ and $\nu_3$ lines. As expected, the splittings depend on the perpendicular magnetic field component $B_\perp$. This dependence is quantitatively analyzed in Fig.~\ref{fig_turning}a where the ODMR spectra in the vicinity of the GSLACs for different $B_\perp$ are compared. It is evident from Fig.~\ref{fig_turning}b that the energy gaps between the levels at both GSLACs increase with $B_\perp$, however with different slopes. Also the splitting at GSLAC-2 is smaller than the splitting at GSLAC-1.
The reason is that the GSLAC-1 emerges because of the coupling between the states with spin projection difference $\Delta m =1$ while GSLAC-2 emerges because of the coupling between the states with $\Delta m =2$. The latter coupling requires the account for the trigonal asymmetry of the silicon vacancy or the second order in $B_{\perp}$ terms and, therefore, is weaker.

\section{Optical properties of silicon vacancies}\label{Sec_optics}

The spin of a vacancy-related center can be efficiently initialized and read-out by optical means. The optical excitation followed by  spin relaxation in the excited state or spin-dependent recombination via a metastable state leads to a preferential population of the $m = \pm 1/ 2$ spin sublevels in the ground state, see Fig. 1b. The intensity of the optical cycle is also spin dependent: for the V2 center in 4H-SiC studied here the intensity of photoluminescence is higher when the system is in the $m = \pm 3/ 2$ states and lower when the system is in the $m = \pm 1/2$ states, which is used for optical read-out of the spin state. The preferential optical pumping of certain spin states together with the spin dependence of the optical cycle rate underlies the microscopic mechanism of ODMR signal formation. It also follows that the PL intensity is sensitive to the variation of spin relaxation rate or to the spin state mixing which occurs at level anticrossings.

\begin{figure}[t]%
\includegraphics*[width=.95\linewidth]{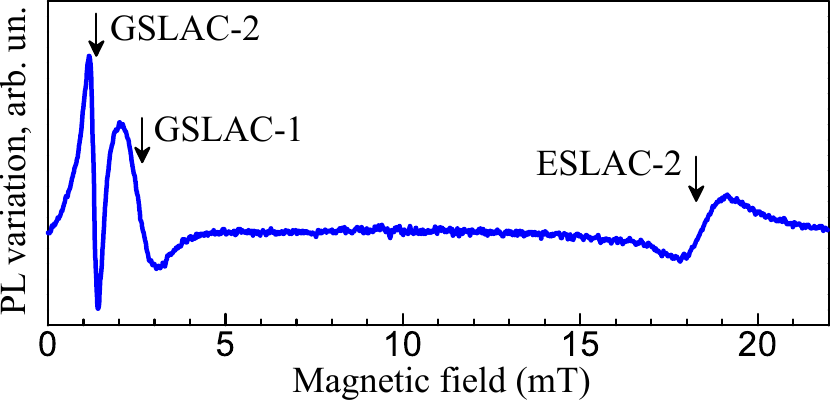}
\caption{Variation of PL intensity with the magnetic field oriented along the $c$-axis. The mixing of spin states at the vicinity of level anticrossings in the ground and excited states leads to a pronounced change of the PL intensity. The data are obtained at $T = 80$~K.}
\label{fig2}
\end{figure}

Strong variation of PL intensity in the vicinity of level anticrossings is indeed observed in experiments.
Figure~\ref{fig2} shows the dependence of the PL intensity on the magnetic field $\bm B$ applied along the $c$-axis. For better visibility, the dependence is measured and presented as the first deviation of the PL signal, which is initialized by a small alternating magnetic field applied additionally to the static one. At the magnetic fields corresponding to the level anticrossings in the ground state, GSLAC-1 and GSLAC-2, and the level anticrossing in the excited state, ESLAC-2, the PL intensity varies. Such a behavior is caused by the mixing of spin sublevels with different spin projections. As expected, the sharpest variation of the PL is observed at the ``forbidden'' anticrossings. Here, the mixing of spin sublevels is weaker and, therefore, the magnetic field range where the mixing plays an important role is smaller. 

The above spin-dependent optical properties including the PL variation at GSLACs can be theoretically described in the formalism of the spin density matrix. An ensemble of spin-3/2 centers is described by the $4 \times 4$ spin density matrix $\rho$ with 15 real linearly independent components which stand for 3 components of the spin dipole $\bm p$, 5 components of the spin quadrupole $\bm d$, and 7 components of the spin octupole $\bm f$. The spin density matrix is conveniently expanded in the basis matrices as follows
\begin{equation}
\rho =  \frac{I}{4} + p_0 \frac{J_z}{\sqrt{5}} + \left( p_1 \frac{J_+}{\sqrt{10}} + {\rm H.c.} \right) + \rho_d + \rho_f \,,
\end{equation}
where
\begin{align}
\rho_d  & =  \frac{d_0}{2} \left( J_z^2 - \frac54 \right) +\left( d_{1} \frac{\{ J_z, J_+\}}{\sqrt{6}}  + d_{2} \frac{J_+^2}{\sqrt{24}}     + {\rm H.c.}\right) , \nonumber \\
\rho_f & = f_{0} \, \frac{\sqrt5}3 \left( J_z^3 - \frac{41}{20}J_z\right) + \left( f_{1}\, \sqrt{\frac5{12}} \left\{ J_z^2-\frac{17}{20}, J_+  \right\} \right.\nonumber \\
&  \left. +  \frac{f_2}{\sqrt{6}} \{ J_z, J_+^2\} + \frac{f_3}{6} J_+^3 + {\rm H.c.} \right).
 \nonumber
\end{align}
In these notations, the components $p_0$, $d_0$, and $f_0$ are real while the other dipole, quadrupole and octupole components have complex values. The components of the average spin $\bm s$ are related to the components of the spin dipole by $s_z = \sqrt{5} \, p_0$, 
$s_x = \sqrt{10} \, {\rm Re} \, p_1$, and $s_y = - \sqrt{10} \, {\rm Im} \, p_1$.

Of particular interest for our study is the quadrupole component $d_0$. It describes the difference between the populations of the spin states $m = \pm 3/2$ and the states $m = \pm 1/2$. This difference determines the optical cycle rate and, hence, affects the PL intensity. The corresponding spin-dependent correction to the PL intensity $I_{\rm PL}$ can be described by
\begin{equation}
\Delta I _{\rm PL}= \alpha d_{0} I_{\rm PL} \,,
\end{equation}
where $\alpha$ is a positive parameter. PL intensity variation is about $0.5\%$ when $d_0=0.05$, which provides the estimation $\alpha \sim 0.1$. 

In thermal equilibrium the spin density matrix is diagonal and all the component of $\bm p$, $\bm d$, and $\bm f$ are vanishingly small since the thermal energy $k_B T$ is much larger than the spin splittings in the ground state. 

Optical excitation of vacancies creates a non-zero component $d_0$ and, therefore, modifies the PL intensity. For a cw radiation, the steady-state spin density matrix is found from the quantum kinetic equation
\begin{equation}\label{rho_eq}
\frac{i}{\hbar} [{\cal H}, \rho] = G + (\dot{\rho})_{\rm rel} \,,
\end{equation}
where $G$ is the generation matrix and $(\dot{\rho})_{\rm rel}$ is the term describing the processes of spin relaxation. 

The optical pumping leading a preferable population of the $m = \pm 1/2$ spin states is given by the generation matrix
\begin{equation}
G = -  \frac{\eta}{2} \left(J_z^2 - \frac54 \right) I_0 \,,
\end{equation}
where $\eta$ is a positive parameter and $I_0$ is the intensity of incident radiation. 
The spin relaxation in the ensemble of spin-3/2 centers in the isotropic approximation
is described by three characteristic times: the relaxation times of the spin dipole $T_p$, spin quadrupole $T_d$, and spin octupole $T_f$.

At zero external magnetic field, $d_0$ is decoupled from the other components of the spin density matrix, 
and the solution of Eq.~\eqref{rho_eq} is straightforward: $d_0 = - \eta T_d I_0$. It corresponds to the PL intensity correction
\begin{equation}
\Delta I_{\rm PL} = - \alpha \eta T_d I_0 I_{\rm PL}
\end{equation}
which is proportional to the relaxation time of the spin quadrupole $T_d$ and scales quadratically with the intensity of incident radiation $I_0$. Excitation with the intensity $I_0 = 10\,$W/cm$^2$ leads to $\sim 10\%$ spin alignment at room temperature,
when $T_d \sim 300 \,\mu$s, which allows us to estimate that $\eta \sim 10^{-6}\,$cm$^2/$erg. 

In the vicinity of an anticrossing, the spin states are mixed. For example, at GSLAC-1, the effective Hamiltonian~\eqref{H} rewritten in the matrix form is given by 
\begin{equation}
{\cal H} = \left( 
\begin{array}{cccc}
E_{3/2} & 0 & 0 & 0 \\
0 & E_{1/2} & 0 & 0 \\
0 & 0 & E_{-1/2} & \Lambda_1/2 \\
0 & 0 & \Lambda_1/2 & E_{-3/2}
\end{array}
\right) ,
\end{equation}
where the energies $E_{-1/2}$ and $E_{-3/2}$ are close to each other and $\Lambda_1$ is the level splitting at GSLAC-1. 

In the relevant case of long relaxation times, when $T_p,T_d,T_f \gg 1/\Lambda_1$, only the multipoles with the zero angular momentum projection $p_0$, $d_0$, and $f_0$ are significant. The other multipoles are small because of rapid oscillations in the magnetic field.
The iterative consideration of Eq.~\eqref{rho_eq} in the vicinity of GSLAC-1 yields the matrix equation for $p_0$, $d_0$, and $f_0$
\begin{align}\label{eq3x3}
\bm R_1  \left( \begin{array}{c} p_0 \\ d_0 \\ f_0 \end{array}\right)  + \left( \begin{array}{c} p_0/T_p \\ d_0/T_d \\ f_0/T_f \end{array}\right)= \left( \begin{array}{c} 0 \\ - \eta I_0 \\ 0 \end{array}\right) , 
\end{align}
where $\bm R_1$ is the matrix describing the multipole mixing,
\begin{align}
 \bm R_1 &= \frac{\Lambda_1^2 (3/T_p+5/T_d+2/T_f)}{100(E_{-1/2}-E_{-3/2})^2}  
\left( 
\begin{array}{ccc} 
 1 & -\sqrt5 & 2 \\
 -\sqrt5 & 5 & -2\sqrt5 \\
 2 & -2\sqrt5 & 4
 \end{array}
\right) .
\end{align}

By solving Eq.~\eqref{eq3x3} we obtain the steady-state values of the spin multipoles $s_0$, $d_0$, and $f_0$ and, hence,
the modification of the PL intensity. Such a calculation gives the following dependence of the PL intensity on the magnetic field $B_z$ in the vicinity of GSLAC-1 
\begin{align}\label{dI}
\frac{\Delta I_{\rm PL}}{I_{\rm PL}} = - \alpha \eta T_d I_0 \left[ 1 - \frac{A_1 \delta_1^2}{(B_z-B_{1})^2 + \delta_1^2}\right] ,
\end{align}
where $B_1 = 2 D /g\mu_B$ is the magnetic field at which GSLAC-1 occurs, 
\begin{align}
&A_1 = \frac{5T_d}{5T_d+4T_f+T_p} \,,\\\nonumber
&\delta_1^2 = \left(\frac{\Lambda_1}{g\mu_B}\right)^2 \frac{(5T_d+4T_f+T_p)(5/T_d+2/T_f+3/T_p)}{100} \,.
\end{align}
The peak in the dependence $\Delta I (B_z)$ has the Lorentz shape with the amplitude $A_1$ and the width $\delta_1$.

In addition to the change in the PL intensity, the mixing of spin states at GSLAC-1  leads to the emergence of the average spin $s_z$ and the spin octupole component $f_0$,
\begin{align}\label{sz}
s_z &= -   \frac{\eta \, T_p I_0 \, A_1 \delta_1^2}{(B_z-B_1)^2 + \delta_1^2} \,, \;\; f_0 = \frac{2}{\sqrt{5}} \frac{T_f}{T_p} s_z \,.
\end{align}
We emphasize that the spin polarization of V$_{\rm Si}$ is induced here by linearly polarized (or unpolarized) radiation. Moreover, the optically induced spin polarization can easily exceeds the vanishingly small equilibrium polarization determined by the ratio $g \mu_B B/k_B T$ which is $\sim 10^{-5}$ for the magnetic field $B = 3$~mT and the temperature $T = 300$~K. The emergence of spin polarization can be observed, e.g., via the spin Faraday or spin Kerr effects.

Similar analysis and calculations can be carried out for the mixing of the states $m = +1/2$ and $m = -3/2$ at GSLAC-2 which occurs in the magnetic field $B_2 = D/(g\mu_B)$. In this case, the steady-state spin multipoles satisfy Eq.~\eqref{eq3x3} with the matrix of the multipole mixing
\begin{align}
 \bm R_2 &= \frac{\Lambda_2^2 \, (1/T_d+1/T_f) }{20(E_{+1/2}-E_{-3/2})^2} 
\left( 
\begin{array}{ccc} 
 4 & -2\sqrt5 & -2 \\
 -2\sqrt5 & 5 & \hphantom{2}\sqrt5 \hphantom{-}\\
 -2 & \sqrt5 & 1
 \end{array}\right) ,
\end{align}
where $\Lambda_2$ is the level splitting at GSLAC-2. The PL variation at GSLAC-2 is then given by
\begin{align}\label{dI2}
\frac{\Delta I_{\rm PL}}{I_{\rm PL}} = - \alpha \eta T_d I_0 \left[1 - \frac{A_2 \, \delta_2^2}{(B_z-B_{2})^2 + \delta_2^2}\right] ,
\end{align}
where
\begin{align}
&A_2 = \frac{5T_d}{5T_d+T_f+4T_p} \,,\\\nonumber
&\delta_2^2 = \left(\frac{\Lambda_2}{g\mu_B}\right)^2 \frac{(5T_d+T_f+4T_p)(1/T_d+1/T_f)}{80} \,.
\end{align}
The average spin $s_z$ and the spin octupole component $f_0$ also emerge at GSLAC-2. They are given by
\begin{align}\label{sz2}
s_z &= - \frac{2 \eta \, T_p I_0 \, A_2 \delta_2^2}{(B_z-B_2)^2 + \delta_2^2} \,, \;\; f_0 = - \frac{1}{2\sqrt{5}} \frac{T_f}{T_p} s_z \,.
\end{align}
Since the level splitting at GSLAC-2 is smaller than the level splitting at GSLAC-1 (see the discussion above), the PL intensity variation is more pronounced at GSLAC-2, which is in accordance with the experiment.

\section{All-optical magnetometry and thermometry}\label{Sec_metry}

The fact that the PL intensity varies in the vicinity of level anticrossings, as it was demonstrated experimentally and explained theoretically in Sec.~\ref{Sec_optics}, suggests a method for all-optical sensing of the magnetic field. The experimental procedure is straightforward and does not require the application of an additional RF magnetic field, which is typically used in high sensitive magnetometry based on color centers. We tune our system to GSLAC-2, characterized by the narrowest resonance (see Fig.~\ref{fig2}) and monitor the PL intensity through the lock-in signal. Small variations of the magnetic field along the $c$-axis is applied to induce a change in the PL intensity. 

\begin{figure}[t]%
\includegraphics*[width=.99\linewidth]{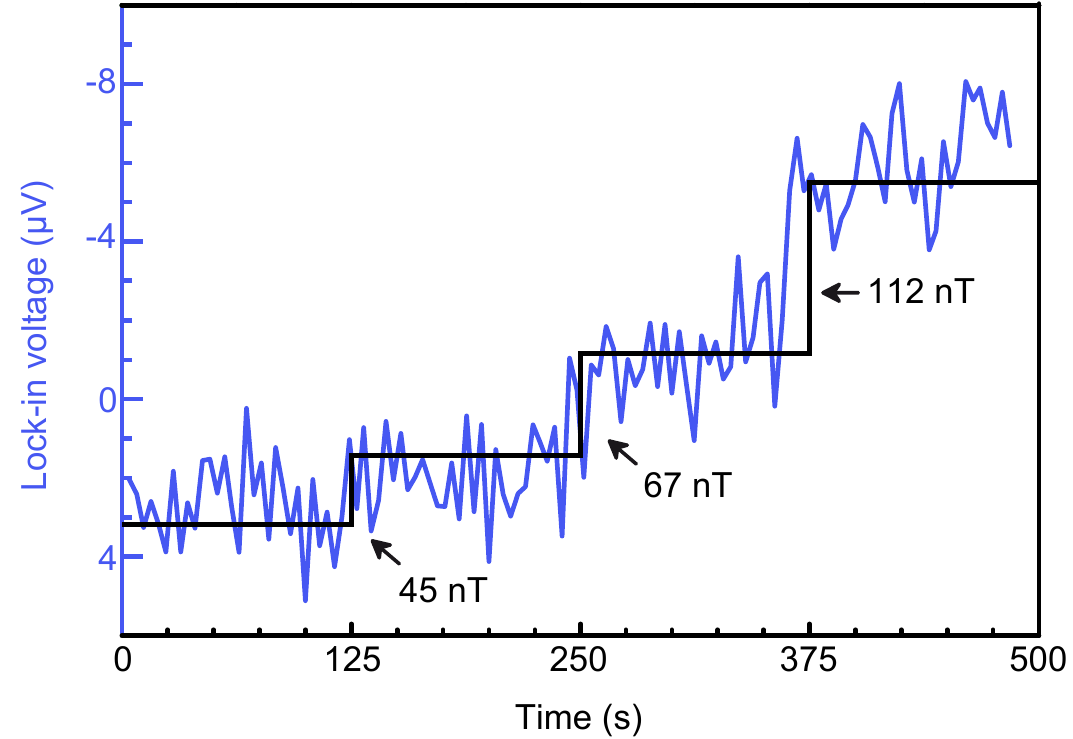}
\caption{Response of the PL signal, measured as the lock-in in-phase voltage, to the magnetic field increase by $45$, $67$, 
and $112\,\mathrm{nT}$ starting from the magnetic field of GSLAC-2.}
\label{fig_steps}
\end{figure}

Figure~\ref{fig_steps} shows the time dependence of the PL signal for the magnetic field changing by steps starting from the magnetic field $B_2$ at GSLAC-2. The PL intensity exhibits clearly pronounced changes. The analysis of the signal-to-noise ratio shows that one 
achieves the dc magnetic field sensitivity better than 100 nT$/\sqrt{\rm Hz}$ within a volume of 3$\times$10$^{-7}$~mm$^3$~\cite{Simin2016}. The approach does not require the application of RF fields and, therefore, is scalable to much larger volumes. For an optimized light-trapping waveguide of 3~mm$^3$, the projection noise limit is below 100 fT$/\sqrt{\rm Hz}$.

\begin{figure}[t]%
\includegraphics*[width=.95\linewidth]{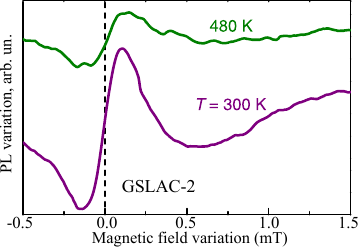}
\caption{Variation of PL intensity with the magnetic field oriented along the $c$-axis at the vicinity of GSLAC-2. Curves are measured at different temperatures.}
\label{fig3}
\end{figure}

The zoom-in of the PL intensity variation with the magnetic field in the vicinity of GSLAC-2 is shown in Fig.~\ref{fig3}.
The figure reveals that the used method of magnetometry is robust at high temperatures up to at least 500 K, suggesting a simple, contactless method to monitor weak magnetic fields in a broad temperature range. While this method is limited to detection of only one component of the magnetic field, vector magnetometry schemes measuring also the field  direction using $V_{\rm Si}$ color centers in SiC were proposed as well~\cite{Lee2015,Simin2015}.

\begin{figure}[t]%
\includegraphics*[width=.99\linewidth]{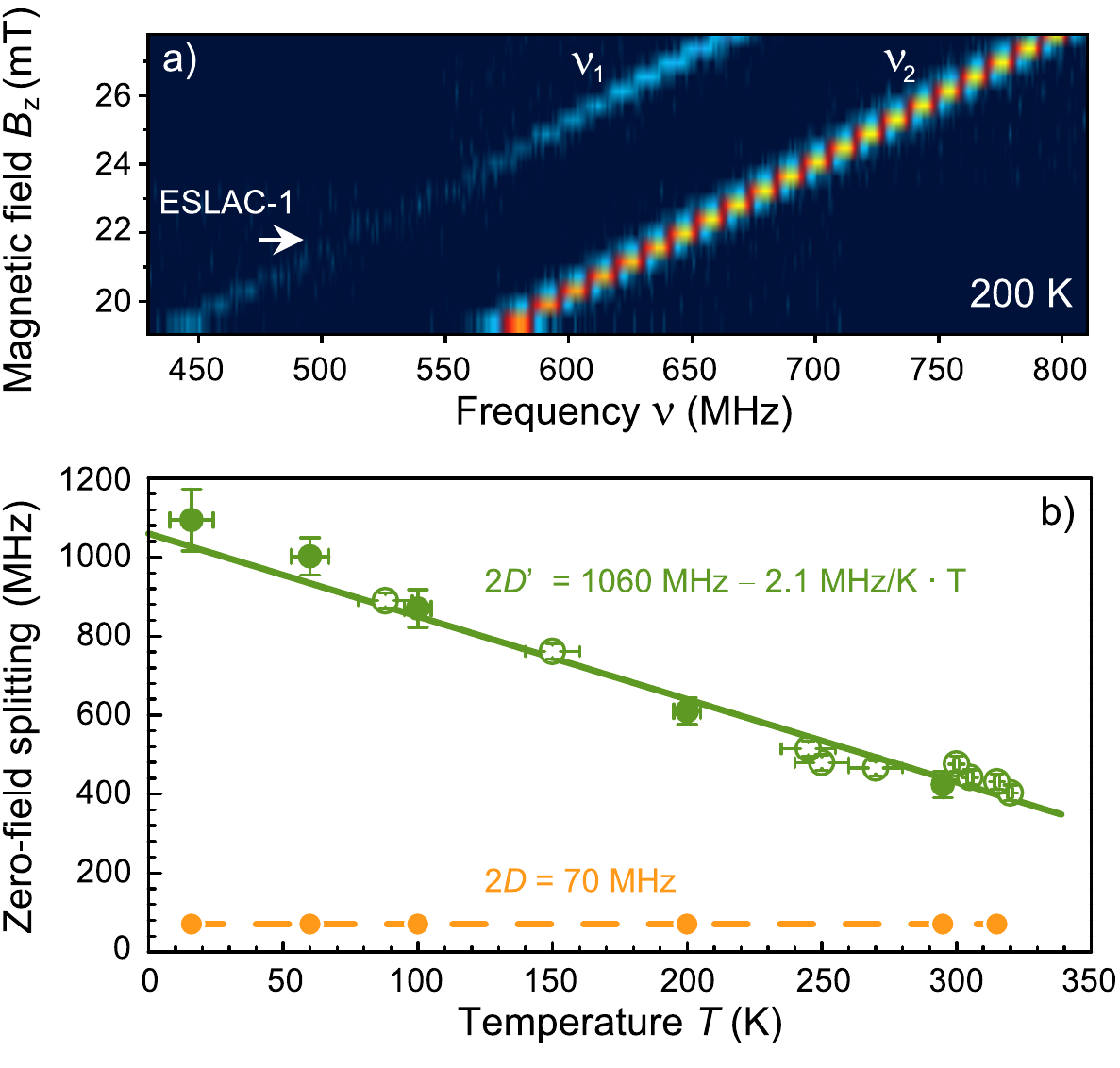}
\caption{(a) Evolution of the $\nu_1$ and $\nu_2$ ODMR lines in the vicinity of ESLAC-1 at $T = 200 \mathrm{K}$. From the change in the contrast of the $\nu_1$-line, the position of ESLAC-1 and hence the zero-field splitting of the excited state $2D'$ can be determined. 
(b) Thermal dependence of the zero-field spin splittings in the ground and excited states obtained from the ODMR measurements (full circles) and with the all-optical technique (empty circles).}
\label{fig4}
\end{figure}

The positions of level anticrossings in the magnetic field for the ground and excited states are determined by the corresponding constants of zero-field spin splitting. The zero-field splittings depend, in turn, on temperature or strain, which can be used for measurements of temperature or strain applied to the structure. In particular, it is found that while the zero-field spin splitting in the ground state $2D$ is almost temperature independent, the zero-field splitting in the excited state $2D'$ has a giant thermal shift of about $2.1$~MHz/K~\cite{Anisimov:2016er}. The results are summarized in Fig.~\ref{fig4}b. 

The thermal shift of the zero-field splitting in the excited state can be deduced from the ODMR measurements by monitoring the magnetic field position corresponding to the change in the $\nu_1$ line contrast which occurs at ESLAC-1. At room temperature, ESLAC-1 occurs at around $15\,$mT, as seen in Fig.~\ref{fig_ODMR}b. The results of the measurements at $T=200\,\mathrm{K}$ depicted in Fig.~\ref{fig4}a show that the contrast changes and, accordingly, ESLAC-1 occur at around $22\, \mathrm{mT}$. Alternatively, the zero-field splitting in the excited state can be determined RF-free way by monitoring the PL intensity in the vicinity of the excited-state level anticrossing ESLAC-2. The dependence of $2D'$ on temperature determined from both techniques together with a linear fit are depicted in 
Fig.~\ref{fig4}b. The observed strong temperature shift of the ESLAC position suggests an all-optical thermometry. Here, we achieve a sensitivity better than 100~mK$/\sqrt{\rm Hz}$ with the volume 10$^{-6}$~mm$^3$ and the projection limit below 100~mK$/\sqrt{\rm Hz}$ for the volume 1~mm$^3$~\cite{Anisimov:2016er}.

We note that the sensing techniques described above utilize different level anticrossings. The all-optical magnetometry scheme relies on GSLAC-2, which shows no temperature shift, while the thermometry relies on the giant temperature shift of ESLAC-2.
It suggests that all-optical integrated magnetic field and temperature sensors can be implemented on the same $V_{\rm Si}$ center.

\section{Spin fluctuations and optical blinking}\label{Sec_noise}

The spin and optical properties of color centers can be also studied by probing the fluctuations stemming from the interaction of the centers with environment. Owing to the fundamental connection between fluctuations and dissipation processes~\cite{Callen1951}, such a noise spectroscopy is a powerful tool for studying the spin dynamics in conditions close to thermal equilibrium and beyond, which has been already demonstrated for atomic systems and semiconductors~\cite{Zapasskii2013review,Hubner2014}. 

Sensitivity of the optical cycle to the spin state of a vacancy implies that the intensity of PL from an individual center exhibits
fluctuations (blinking) caused by the jumps of the spin center between the ``bright'' states ($m = \pm 3/2$) and the ``dark'' states ($m = \pm 1/2$). Our analysis shows that, for an ensemble of vacancies in zero magnetic field, the spectral density of the PL intensity fluctuations $\langle \Delta I_{\omega}^2 \rangle = \int \langle \Delta I (0) \Delta I (t) \rangle {\rm e}^{{\rm i} \omega t} dt$ is given by the Lorentz function
\begin{equation}
\langle \Delta I_{\omega}^2 \rangle \propto \dfrac{1}{1+(\omega T_d)^2} \,,
\end{equation}
where the angle brackets denote averaging over the vacancies. It is naturally expected that the time $T_d$ depends on temperature and the intensity of incident radiation since the relaxation may be much more efficient when the centers are in excited states. 

The maximum of the spectral density of the PL intensity fluctuations can be shifted from zero frequency by applying a RF field inducing the resonant spin transitions between the spin sublevels $\pm 3/2$ and $\pm 1/2$ which are split by $2D$. In this case, the spectral density of the PL intensity fluctuations assumes the form
\begin{equation}
\langle \Delta I_{\omega}^2 \rangle \propto \dfrac{1}{1+(\omega - \Omega_R)^2 T_d^{2}} \,,
\end{equation}
where $\Omega_R$ is the frequency of the Rabi oscillations induced by the RF field. Thus, the spin fluctuations in an ensemble of Si-vacancies can be probed by measuring the correlation function of the luminescence intensity $g^{(2)}$. The spectral density of the PL intensity fluctuations provides information about the spin relaxation time. 

\section{Summary}\label{Sec_summary}

We have considered the spin structure, spin dynamics, and optical properties of spin-3/2 color centers associated with Si vacancies in 4H-SiC. A sharp variation of the vacancy-related photoluminescence at the vicinity of spin level anticrossings, which occur at magnetic fields of a few mT in SiC, can be utilized for all-optical sensing of physical quantities such as magnetic field, temperature, strain, etc. We have presented experimental data on all-optical magnetometry and thermometry and developed a microscopic theory of the optical signal variation at level anticrossings. The theory describes the observations and predicts the emergence of a strong non-equilibrium spin polarization of vacancy-related centers at level anticrossings. We also have discussed the spin noise in an ensemble of spin-3/2 centers and proposed a way to measure the spin relaxation time by analyzing the correlation function of the photoluminescence intensity.

\paragraph*{Acknowledgments.} 
This work has been supported by the Russian Science Foundation grant No. 16-42-01098,
the German Research Foundation (DFG) grants DY 18/13 and AS 310/5,  the ERA.Net RUS Plus program, the German Federal Ministry of Education and Research (BMBF) within the project DIABASE, and the Foundation ``BASIS''.



%

\end{document}